# Bed load proppant transport during slickwater hydraulic fracturing: insights from comparisons between published laboratory data and correlations for sediment and pipeline slurry transport

Mark McClure

McClure Geomechanics LLC

mark@mccluregeomechanics.com

November 12, 2016

## Abstract

Bed load transport is the movement of particles along the top of a bed through rolling, saltation, and suspension created by turbulent lift above the bed surface. In recent years, there has been a resurgence of interest in the idea that bed load transport is significant for proppant transport during hydraulic fracturing. However, scaling arguments suggest that bed load transport is only dominant in the laboratory and is negligible at the field scale. This paper revisits the discussion and provides new analysis.

This paper focuses on bed load transport in thin fluid such as water. In slickwater fracturing, injection rate is high, proppant concentration is low, and particle settling is relatively rapid. As a result, slickwater fracturing is the type of hydraulic fracturing where bed load transport is most likely to be significant.

I review laboratory experiments that have been used to develop concepts of bed load transport in hydraulic fracturing. I also review the scaling arguments and laboratory results that have been used to argue that viscous drag, not bed load transport, is dominant at the field scale.

I compare literature correlations for fluvial sediment transport and for pipeline slurry transport with published laboratory data on proppant transport in slot flow. The comparisons indicate that fluvial transport correlations are suitable for predicting the rate of sediment erosion due to flow of proppant-free fluid over a bed. The pipeline slurry correlations are suitable for predicting the ability of proppant that is already in suspension to flow in bed transport without settling, but only if the aspect ratio of the flowing region in the slot is close to unity. This can occur at the laboratory scale, but not in the field.

The comparison indicates that at low rates of proppant flow, the "equilibrium bed height" (or equivalently, the equilibrium superficial velocity) in a laboratory proppant transport experiment can be predicted from pipeline slurry correlations. However, as the volumetric flow rate and the aspect ratio of the flowing region in the slot increase, the applicability of the pipeline slurry correlations breaks down. This demonstrates that the scaling of proppant transport in low rate experiments is different from the scaling in high rate experiments. Results from low rate experiments cannot be directly extrapolated to the field scale. Laboratory experiments using higher volumetric flow rate indicate that bed load transport rates are too low to be significant at the field scale, even if the flow velocity is very high. Based on the laboratory results, I propose an equation to estimate the maximum possible rate of bed load proppant transport at the lab and field scale. Overall, the results indicate that bed load transport is a negligible process at the field scale, except under a narrow set of unusual circumstances.

## 1. Introduction

Bed load transport occurs when solid particles are transported along a surface by rolling, saltation, and suspension. Bed load transport has been observed to be a dominant process in laboratory studies of proppant transport through a slot and has been proposed as a dominant process for proppant transport in the field (Kern et al., 1959; Patankar et al., 2002; Wang et al., 2003; Brannon et al., 2005; Woodworth and Miskimins, 2007; Mack et al., 2014). In spite of these observations, some authors have argued that bed load transport is not a significant process at the field scale (Biot and Medlin, 1985; Medlin et al., 1985). They pointed out that the rate of bed transport does not increase as fracture height increases. On the other hand, the particle settling distance, which scales with convective transport from viscous drag, does increase with fracture height. As a result, they argued that bed transport is a dominant process at the laboratory scale but plays a minor role in field scale fracturing. Conversely, they argued that convective transport from viscous drag is negligible in the laboratory scale but dominant in the field. In this context, the word "convection" is used to refer to proppant being carried by viscous drag from the horizontal movement of the flowing fluid, not to refer to gravitationally driven slurry convection (Clark, 2006).

In recent years, there has been a resurgence of interest in bed load transport, apparently because of the widespread adoption of slickwater fracturing. With low viscosity fluids, proppant settling velocity is high, and so viscous drag is less effective as a mechanism of proppant transport. Industry publications are increasingly adopting the view that bed load transport is a dominant process in slickwater fracturing (Patankar et al., 2002; Wang et al., 2003; Brannon et al., 2005; Woodworth and Miskimins, 2007; Mack et al., 2014). Correlations based on bed load transport have been incorporated into field scale fracturing simulators (Weng et al., 2011; Shiozawa and McClure, 2016). This paper revisits the issue of whether bed load transport plays a significant role at the field scale.

I review several laboratory experiments involving proppant flow through a slot. Figure 1 shows a schematic of these experiments. Fluid/proppant slurry is injected from left to right. The proppant settles due to gravity and forms into an immobile bed at the bottom of the slot. At the far left, turbulence stimulated by flow from the inlet creates additional lift and the immobile proppant bed is relatively short (Medlin et al., 1985; Patankar et al., 2002; Brannon et al., 2005). Moving away from the inlet, there is a viscous drag region where particles tend to settle downward as they are carried by the flowing fluid through the slot. At a sufficient distance from the inlet, the thickness of the viscous drag region goes to zero, and particle transport is dominated by bed load transport. If the height of the bed is less than the "equilibrium bed height," proppant deposition occurs over time. As the bed height grows, the cross-sectional area available for flow reduces, increasing the flow velocity, which increases the ability of proppant to be transported through the slot without deposition. Eventually, the equilibrium bed height (and corresponding equilibrium velocity) is reached, and height growth ceases. Conversely, if the height is greater than the equilibrium height, then flow velocity is greater than the equilibrium velocity, and there is net erosion of proppant. The height reduces over time, reducing flow velocity, until equilibrium is reached.

The height of the region flowing above the bed is $H_1$. The height of the bed load transport region is $H_b$, and the distance from the top of the bed load transport region to the top of the slot is $H_2$. The volumetric flow rate of proppant and fluid is $Q_t$. The superficial velocity of the slurry, $V_{avg}$, is equal to $Q_t/(H_1 W)$, where $W$ is the slot width.

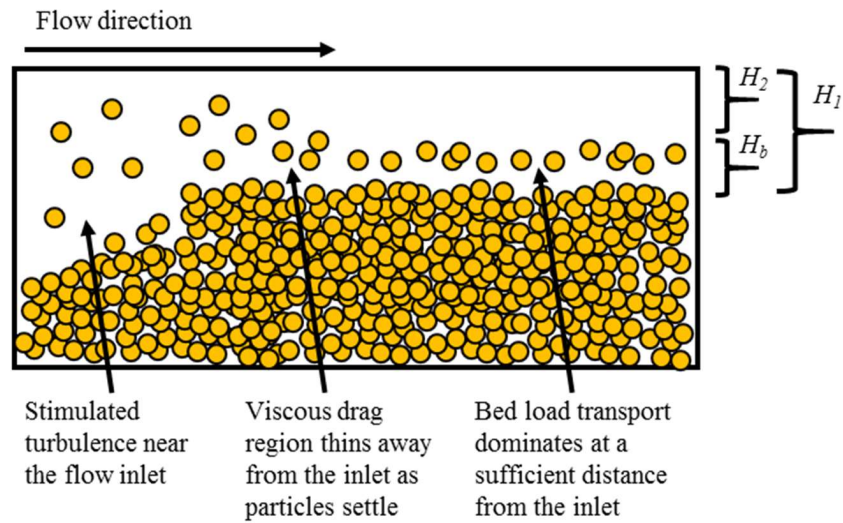

Figure 1: Schematic of a slot flow proppant transport laboratory experiment

### 1.1 Literature supporting the bed load transport concept

Kern et al. (1959) performed the earliest experimental work on proppant transport in hydraulic fracturing. They pumped fluid/sand slurry into a slot that was 0.25 m tall, 0.56 m long, and 0.00635 m wide. They reported their results in a plot of equilibrium velocity versus mass rate of proppant injection. Kern et al. (1959) used these experiments to propose a mechanism for proppant transport during hydraulic fracturing. At the beginning of the treatment, the sand is transported horizontally by flow of the injection fluid (viscous drag) and downward by gravity, settling into a bed. The increasing bed height increases linear velocity until bed load transport becomes the dominant process and injected sand is "washed over the top of the settled sand bed" to the leading edge of the bank. A more detailed mathematical description of this process was provided by Schols and Visser (1974).

Babcock et al. (1967) and Alderman and Wendorff (1970) performed laboratory experiments in which they observed equilibrium bed height behavior. They developed correlations to predict the critical velocity at equilibrium. These investigators used water in some experiments, but most experiments were performed with viscous gel.

Patankar et al. (2002) performed slot flow experiments in a Plexiglass slot with height 0.305 m, length 2.44 m, and width 0.008 m. First, they performed experiments in which proppant-free water was injected into a slot that already contained an immobile proppant bed. The fluid injection caused proppant to be eroded from the bed. As the proppant bed was eroded from the slot, the cross-sectional area for flow increased, reducing superficial velocity. Eventually, the velocity decreased sufficiently that erosion ceased.

Next, they performed experiments in which fluid/proppant slurry was injected continuously into the slot. At different combinations of flow rate, injected proppant concentration, fluid properties, and proppant properties, they measured the thickness of the immobile proppant bed and the thickness of the region above the proppant bed where bed transport was taking place. On the basis of their measurements, they proposed correlations to predict the equilibrium bed height and the thickness of the bed transport region. Wang et al. (2003) developed additional correlations on the basis of these experiments.

Woodworth and Miskimins (2007) applied the Wang et al. (2003) correlations directly to field scale hydraulic fracture design. They asserted that in slickwater fracturing, "proppant falls from suspension and builds a proppant mound before any form of proppant transport takes place." In other words, they asserted that convective transport through viscous drag of proppant is negligible and that bed load transport is the only important mechanism of proppant transport in slickwater fracturing. Based on this premise, they proposed to design hydraulic fracturing treatments by directly applying the Wang et al. (2003) correlations.

Brannon et al. (2005) performed slot flow experiments in a cell with height 0.56 m, length 4.88 m, and width of 0.0127 m or 0.00645 m. They observed an equilibrium bed height develop at a variety of different combinations of flow rate and proppant and fluid properties.

Mack et al. (2014) reviewed a variety of bed load transport processes. They used a correlation based on the Shields number to estimate the critical flow rate required to initiate proppant movement, and then proposed to use this correlation for field scale fracturing design. They implicitly assumed that if threshold conditions are reached such that bed load transport can initiate, then proppant transport will be a significant process.

*1.2 Literature questioning that bed load transport is relevant at field scale*

In spite of the considerable literature on the bed load transport concept for hydraulic fracturing, Medlin et al. (1985) and Biot and Medlin (1985) argued that the observed dominance of bed load transport was a laboratory artifact. Medlin et al. (1985) stated "bed load transport … is unimportant on the scale of field treatments because it does not scale up with slot height."

Medlin et al. (1985) performed experiments in a slot with height 0.305 m, length 6.1 m, and width 0.0095 m. They used an optical absorption method to measure sand density along a vertical section during flow. In the immobile bed, the volumetric fraction of proppant was around 0.5. In the bed transport region, the concentration had a sharp gradient, dropping from 0.5 to about 0 to 0.05 at the top of the bed load transport region. Above the bed transport region, the proppant concentration was low. A viscous drag region was visible above the bed transport region at observations points near the inlet.

Medlin et al. (1985) found that the height of the bed load transport region increased with slurry concentration, viscosity, and volumetric flow rate, but the thickness of the bed load transport region was not affected by the height of the nearly proppant free region above it. They concluded that "viscous drag is the only observable transport mechanism which scales up with slot height … viscous drag accounts for essentially all of the sand transport in large, thin-fluid fracturing treatments." Biot and Medlin (1985) developed analytical expressions, drawing on the theory of sediment transport, to theoretically explain the experimental results from Medlin et al. (1985).

*1.3 Scaling*

In this section, simple scaling arguments are used to explain differences between proppant transport in the field and in the laboratory.

The height that a particle of proppant must fall to settle on the bed can be defined as $H_s$. A vertical fracture propagating from a horizontal well will extend both upward and downward, with a general tendency to extend upward because of the vertical trend in stress with depth. This will cause upward and downward fluid flow from the well. Downward fluid flow will accelerate the settling of the proppant to the bottom of the fracture, while upward fluid flow will pull the proppant upward and delay the settling of the proppant to the bottom of the fracture. Therefore, different proppant grains will have different effective values of $H_s$, with a maximum value equal to the entire height of the fracture (particles carried upward by fluid reaching the upper crack tip) and a minimum value of zero (particles carried downward by fluid reaching the lower crack tip). The "average" value of $H_s$ can be taken as half the total fracture height, and the "maximum" value of $H_s$ can be taken as the total height of the fracture.

The settling velocity, $V_t$, can be calculated from Equation 1 (Ferguson and Church, 2006). This correlation for settling velocity is convenient because it is valid for laminar, transitional, and turbulent flow:

$$V_t = \frac{Rgd^2}{\frac{18}{\rho_f} + \sqrt{0.75Rgd^3}},$$

(1)

where $R$ is equal to $(\rho_s - \rho_f)/\rho_f$, $\rho_s$ is the density of the grains, $\rho_f$ is the density of the fluid, $\mu$ is the fluid viscosity, $g$ is the gravitational constant, and $d$ is the particle diameter.

Stokes law can be used to predict settling rates but is only valid for low particle Reynolds numbers. At high particle Reynolds number, turbulent drag greatly reduces settling rate. For example, Stokes law overestimates the settling rate of 20 mesh proppant in 0.3 cp fluid (the viscosity of water at 80°C, which is a typical reservoir temperature) by a factor of 17. At conditions typical for slickwater fracturing, Equation 1 predicts that $V_t$ is around 0.12 m/s for 20 mesh proppant, 0.08 m/s for 40 mesh proppant, and 0.03 m/s for 100 mesh proppant.

Clustered settling can accelerate settling velocity at volume fractions below 0.1, but only in quiescent or slowly moving fluid (Kirby and Rockefeller, 1985; McMechan and Shah, 1991; Brannon et al., 2005; Liu and Sharma, 2005). Conversely, hindered settling can slow the rate of settling at high concentrations (McMechan and Shah, 1991). The horizontal superficial velocity can be defined as $V_h$. The proppant can travel horizontally through the fracture at a velocity slightly greater than the superficial velocity of the slurry because of particle migration away from the fractures walls (Barree and Conway, 1995). For simplicity, these processes are neglected in the present discussion.

The distance that a particle of proppant can travel before settling, $L_s$, is equal to $V_h H_s / V_t$. In laboratory experiments, $H_s$ may range from 0.01 to 0.3 m, and $V_h$ can range from 0.5 to 5.0 m/s. Therefore, with $V_t$ in the range of 0.1 m/s, the value of $L_s$ in the lab will be no more than 1.5 m and may be much less. In the field, $H_s$ may be in the range of 5-100 m, depending on height growth. Assuming $V_h$ equal to 0.3 m/s and $H_s$ equal to 30 m, an estimate for $L_s$ in the field is 90 m. These calculations demonstrate that convective transport through viscous drag is much greater in the field than in the lab. They also demonstrate that viscous drag is capable of transporting large mesh proppant a reasonable distance away from the well into the formation during slickwater fracturing.

The volumetric rate of proppant transport from viscous drag is equal to the total volumetric flow rate, $Q_t$, multiplied by the proppant volumetric fraction, $C$. The proppant mass flow rate scales directly with injection rate, which is much higher in the field than in the lab. Therefore, far greater mass transport rates and distances are possible from viscous drag in the field than in the lab. If injecting 0.265 m³/s (100 bpm) at volume fraction of 0.05 (1 ppg), the proppant injection rate is 35 kg/s. Typical laboratory injection rates are several orders of magnitude lower.

The volumetric rate that proppant can move in bed transport can be defined as $Q_b$. Experiments of bed transport in the lab have shown that $Q_b$ is no more than a few 0.1s of kg/s. If the bed transport rate is the same in the lab as in the field, this implies that bed transport accounts for a very small fraction of the total proppant movement taking place.

Medlin et al. (1985) and Biot and Medlin (1985) argued that $Q_b$ does not scale up as fracture height increases. On the other hand, Wang et al. (2003) developed correlations on bed load transport from experiments performed by Patankar et al. (2002). These correlations implicitly assume that proppant transport ability scales with fracture size (Woodworth and Miskimins, 2007).

The key issue to address is how $Q_b$, the rate of bed load proppant transport, scales from the laboratory to the field scale.

## 2. Review of relevant concepts from the theories of fluid dynamics, slurry pipeline transport, and sediment transport

In Section 2.1, I review basic properties of turbulent and laminar flow and calculate conditions under which turbulent flow will occur during slickwater fracturing. In Section 2.2, I review fundamental mechanisms of bed load transport. In Section 2.3, I review correlations for predicting the onset of sediment bed erosion using the Shields number. In Section 2.4, I review correlations for predicting the rate of sediment bed load transport during flow in a river. In Section 2.5, I review correlations that can be used to predict the ability of slurry to flow in a pipe without sediment deposition taking place.

The correlations in Section 2.4 apply to systems where sediment is being eroded from an already-existing bed. The correlations in Section 2.5 apply to systems where sediment is flowing in slurry and has not yet been deposited.

*2.1 Turbulent and laminar flow*

The Reynolds number, *Re*, expresses the ratio of inertial to viscous forces during flow (Middleton and Wilcock, 1994; Bird et al., 2006):

$$Re = \frac{\rho_f V L}{\mu},$$
(2)

where $V$ is flow velocity, $L$ is a characteristic length scale, $\rho$ is fluid density, and $\mu$ is fluid viscosity. At low values of *Re*, the flow is dominated by viscous forces, and flow is laminar. In laminar flow, flow is smooth, and streamlines do not mix. At large values of *Re*, turbulent flow occurs. In turbulent flow, there are irregular and unpredictable motions called eddies. The transition from laminar to turbulent flow occurs at a critical Reynolds number in the range of 2000-4000. The critical Reynolds number for flow in a rectangular duct ranges between 1800 at aspect ratio near 1.5 to 2800 at infinite aspect ratio, corresponding to flow between parallel plates (Hanks and Ruo, 1966). The characteristic length scale for flow through a slot is the hydraulic diameter, $D_h$, equal to two times the cross-sectional area divided by the perimeter. For infinite aspect ratio (flow between parallel plates), the hydraulic diameter is double the width. For flow in a wide, open channel, such as a river, the characteristic length scale is the depth of the channel.

During hydraulic fracturing, turbulent flow is likely near the well because of the high velocity created by flow convergence. Further from the well, the occurrence of turbulent flow depends on job-specific parameters, such as injection rate, net pressure, fluid viscosity, the number of flowing fractures, and fracture height. For example, if fluid is flowing at 0.132 m³/s (50 bpm) into one wing of a biwing fracture, with viscosity of 0.3 cp, it can be roughly estimated (using the PKN fracture geometry assumption) that the critical Reynolds number will be reached as long as the height is less than 314 m, which is virtually certain. On the other hand, if the fluid viscosity is 10 cp, turbulent flow will only occur if the height is less than 10 m. Aperture is highest along the centerline of the fracture, tapering to zero at the top and bottom, and so turbulent flow is most likely along the center of the fracture. The onset of turbulent flow is associated with non-Darcy pressure drop (Fourar et al., 1993).

When turbulent flow occurs in the presence of a solid boundary (such as the bottom of a river bed or a fracture), a boundary layer forms. A boundary layer is defined as a region of flow that is significantly affected by the presence of a solid boundary. Because boundary layers form in turbulent flow, the region of flow disturbance created by a solid boundary is much larger in laminar flow than in turbulent flow. Flow in the boundary layer may be turbulent or laminar. If a turbulent boundary layer develops, there will be an even thinner laminar, viscous sublayer with a very large velocity gradient (Middleton and Wilcock, 1994).

*2.2 Bed load transport processes*

Pye (1994) provided an excellent description of natural sediment transport processes. There are two types of natural sediment transport processes relevant to proppant transport in hydraulic fracturing: fluvial (river and stream) transport and turbidity currents.

Fluvial sediment transport occurs as water erodes particles from the underlying riverbed and sweeps the sediment downstream. The main erosional processes are rolling (creep or reputation), saltation, and suspension. In saltation, individual particles are abruptly lifted into from the bed by Bernoulli lift created by turbulent velocity fluctuations. In suspension, lift created by turbulence enables the particles to remain in the fluid above the bed for an extended period of time.

Chapter 5 from Wasp et al. (1977) describes how turbulence generates lift above a surface. Gravity tends to create a downward concentration gradient as particles settle and accumulate at the surface. Turbulent mixing resists the creation of concentration gradients, which has the net effect of generating lift on the particles.

Turbidity currents occur when a flood or sudden mass transport event such as a landslide causes a rapid influx of sediment into a large body of water, usually the ocean. The density difference, momentum from the initial event, and the slope of the ocean floor causes the sediment to move very rapidly in plumes that can extend over hundreds of kilometers (Chapter 2 from Wasp et al., 1977; Meiburg and Kneller, 2010).

Turbidity currents and suspended fluvial transport both involve particle suspension from lift created by turbulence above a surface. However, a difference is that fluvial transport involves erosion and suspension of particles that have already settled into a bed. In contrast, turbidity currents involve a sudden influx of rapidly flowing sediment that has not yet been deposited. Of the two, turbidity currents are more similar to the process of proppant injection during hydraulic fracturing. As shown in this paper, correlations for predicting rates of fluvial sediment transport significantly underpredict bed load transport rates if applied to laboratory experiments in which proppant slurry is injected into a slot. However, they accurately predict rates of proppant bed erosion if proppant-free water is pumped over a settled proppant bed.

Solid/liquid slurry flow in pipelines is used for long-distance transportation of coal, ore, and other products (Govier and Aziz, 1972; Wasp et al., 1977). The solid is carried in suspension in the pipeline, and the flow rate must be high enough to prevent solid deposition. This process is analogous to natural turbidity currents and proppant injection during hydraulic fracturing in the sense that it involves suspended bed load flow of particles that have not yet settled.

### 2.3 Correlations for predicting the onset of bed load transport

Bed load transport behavior can be quantified with the Shields number, which expresses the ratio of the shear force to the gravitational force on a particle of sediment (Miller et al., 1977):

$$N_{Sh} = \frac{\tau_b}{(\rho_s - \rho_f)gd}, \tag{3}$$

where $\tau_b$ is the shear stress acting on the top of the bed, and $d$ is the particle diameter. The Shields number is used for describing sediment transport in open channels, such as rivers and streams.

The critical Shields number is the value that is required for the onset of particle motion (erosion) from the surface of the bed. The critical Shields number is a function of the boundary Reynolds number, $Re^*$. The boundary Reynolds number is calculated using the shear velocity, $u^*$, instead of the flow velocity, $V$. The shear velocity is equal to the shear stress on the bed scaled by fluid density:

$$u^* = \sqrt{\frac{\tau_b}{\rho_f}}. \tag{4}$$

Shields diagrams show the critical Shields number as a function of $Re^*$. For values of $Re^*$ greater than 1.0, which is generally true of the conditions during slickwater hydraulic fracturing, the critical Shields number (the value at which bed load transport begins) varies between 0.03 and 0.06 (Miller et al., 1977).

For slot flow, the value of $\tau_b$ may be estimated as (Appendix A, Biot and Medlin, 1985):

$$\tau_b = \frac{1}{8} f_D \rho_f V^2, \tag{5}$$

where $f_D$ is the Darcy friction factor (defined as equal to four times the Fanning friction factor).

The Darcy friction factor can be estimated from a chart or a correlation (Chen, 1979). For laminar flow, $f_D$ is equal to $64/Re$. For turbulent flow, $f_D$ is a function of the Reynolds number and the relative roughness, which is equal to the absolute roughness (the root mean squared amplitude of asperities on the surface) divided by the hydraulic diameter. If the relative roughness is greater than about 0.05, the Darcy friction factor is nearly constant during turbulent flow and is equal to approximately 0.0775. During hydraulic fracturing in the field, the relative roughness may exceed 0.05 because the size of the asperities can be on the order of hundreds of microns, and the aperture can be on the order of millimeters.

A reasonable estimate for the roughness of Plexiglass (used in most slot flow laboratory experiments) is 100 microns (Persson et al., 2005). The laboratory experiments reviewed in this paper used width of approximately 0.006 to 0.0095 m, implying relative roughness on the order of 0.005 – 0.008. At this value of relative roughness, the friction factor is about 0.045 at the onset of turbulence flow and decreases to an asymptotic value of around 0.03 at high Reynolds number (greater than 100,000). Most of the experiments reviewed in this paper were performed at turbulent flow, within an order of magnitude of the critical Reynolds number, and so $f_D$ can be reasonably approximated as 0.04.

Equations 3 and 5 can be combined to give an equation predicting the onset of bed erosion:

$$V_c = \sqrt{\frac{8N_{Sh,c}Rgd}{f_D}} \tag{6}$$

### 2.4 Correlations for fluvial bed load transport

A variety of equations for predicting the rate of bed load transport in a stream or river have been developed (Meyer-Peter and Müller, 1948; Yalin, 1977). These equations were developed for systems where fluid flows over a preexisting bed of sediment. Therefore, their application to hydraulic fracturing is limited to cases were proppant-free or very dilute slurry flows over a preexisting bed of proppant (Section 3.2).

Wiberg and Smith (1989) provided a modified form of the Meyer-Peter and Müller (1948) correlation:

$$Q_s = W(d\sqrt{gdR})(9.64N_{Sh}^{0.166})\left(N_{Sh} - N_{Sh,c}\right)^{3/2}, \tag{7}$$

where $Q_s$ is the solid volumetric flow rate, $W$ is the width available for flow (equal to the aperture in the case of a fracture), $N_{Sh,c}$ is the critical Shields number for the onset of bed load transport (the value used in the original Meyer-Peter and Müller correlation was 0.047).

### 2.5 Correlations for transport of heterogeneous suspensions in horizontal pipelines

For slurry flow in horizontal pipelines, correlations are available for predicting the deposition velocity, $V_D$, the velocity required to prevent solid deposition from occurring (Wasp et al., 1977; Govier and Aziz, 1972). These correlations were developed for particles flowing in the liquid phase that have not yet deposited, and so are relevant to the case of proppant injection during hydraulic fracturing. Transport rates achieved through this process are much higher than through the erosion of an existing bed, as given by the correlations in Section 2.4.

Pipeline slurry flow can be classified as homogeneous or heterogeneous, depending on whether there is a significant vertical concentration gradient. The flow will be heterogeneous under the condition (Charles and Steven, 1972):

$$\frac{V_t}{u^*} > 0.13, \tag{8}$$

where $u^*$ can be calculated from Equations 4 and 5.

Using Equations 4 and 5 and assuming $f_D$ equal to 0.04, Equation 8 can be rearranged to state that the critical threshold velocity for homogeneous flow in turbulent flow as:

$$V_{c,homogeneous} \approx 109 V_t. \tag{9}$$

In slickwater fracturing, for very small 100 mesh proppant (diameter of 0.15 mm), the threshold $V_{c,homogeneous}$ is around 3.3 m/s. For 40 mesh proppant, the threshold is around 8.7 m/s, and for 20 mesh proppant, the threshold is around 13 m/s. Therefore, proppant transport during slickwater hydraulic fracturing should almost always be treated as heterogeneous. The Charles and Stevens (1972) correlation was developed for flow in pipes, and so will underestimate $V_{c,homogeneous}$ for flow in fractures, which have much greater aspect ratio.

A simple correlation for the deposition velocity in a pipe is (Durand, 1952):

$$V_D = F_L \sqrt{2gDR}, \tag{10}$$

where $D$ is the pipe diameter, and $F_L$ is a function of particle diameter and volumetric proppant concentration that varies between 0.75 and 1.5.

Equation 10 does not include a scaling with volumetric fraction of solid particles. Thomas (1962) provided a correlation for calculating the effect of proppant concentration:

$$u_D^* = u_0^* (1.0 + 2.8 \left(\frac{V_t}{u_0^*}\right)^{\frac{1}{3}} C^{1/2}), \tag{11}$$

where $u_D^*$ is the critical shear deposition velocity (converted to $V_D$ with Equations 4 and 5), $u_0^*$ is the critical shear deposition velocity for dilute suspensions, and $C$ is the volumetric fraction of particles. Thomas (1962) provided an expression for $u_0^*$, but here, I provide the simpler Wicks (1968) correlation for $V_D$ at dilute concentration:

$$V_{D,0} = 1.87(d/D)^{1/6} \sqrt{2gDR}. \tag{12}$$

Equation 11 was developed for flow in cylindrical pipes. It should not necessarily be expected to apply to flow through a slot. The aspect ratio (height divided by width) of a cylindrical pipe is 1.0, but the aspect ratio of a slot may be much larger than 1.0. Increasing slot height at constant concentration increases the total amount of particle in the slot, making it more difficult to sustain transport without deposition. Even if slot height is increased while holding the amount of particles in the slot fixed (allowing concentration to proportionally decrease as height is increased), the scaling of $V_D$ will change because the proppant will settle to the bottom of the slot, and the vertical concentration profile will be different than in a pipe or slot with aspect ratio close to 1.0. The slurry velocity profile may not be constant along the height of the slot (especially at low aspect ratio), and so increasing aspect ratio away from 1.0 changes the velocity distribution, affecting the flow behavior and bed load proppant transport.

The lift occurring in bed load transport is provided by contact with an underlying surface, and so the total capacity for bed load transport should scale with the integral of the surface area of the underlying surface multiplied by the cosine of the surface angle with horizontal. This integral is equal to diameter for a pipe and width for a slot.

Therefore, if the aspect ratio of the slot is close to 1.0, it may be possible to apply Equation 11 to predict $V_D$, replacing $D$ with $W$. But if the aspect ratio is deviates significantly from 1.0, Equation 11 should not be applicable.

Equation 12 should be applicable to slot flow for sufficiently dilute concentrations (replacing $D$ with $W$). However, the amount of proppant in a slot scales linearly with slot height at constant concentration. Due to settling, a "dilute" proppant concentration in a slot with high aspect ratio could have high proppant concentration at the bottom of the slot. Therefore, the overall concentration required to be sufficiently "dilute" scales inversely with slot height.

## 3. Comparison of equation predictions with published data

In Sections 3.1, 3.2, and 3.3, I compare experimental data from Patankar et al. (2002) to equations reviewed in Sections 2.3, 2.4, and 2.5, respectively. In Section 3.4, I compare the results of Patankar et al. (2002) and Medlin et al. (1985), who used higher fluid and proppant flow rates than Patankar et al. (2002. In Section 3.5, I review the experiments performed by Kern et al. (1959).

### 3.1 Comparison with experimental observations of the onset of proppant erosion from Patankar et al. (2002)

The experimental setup used by Patankar et al. (2002) is described in Section 1.1. In the first set of experiments, proppant-free fluid was injected at specified volumetric rate into the slot, which already contained a settled bed of proppant. The fluid eroded the proppant until equilibrium was reached and erosion ceased. Patankar et al. (2002) tabulated the height of the region between the top of the settled bed and the top of the slot, $H_1$, for a variety of types of proppant and flow rates, and for minor variations of fluid viscosity.

Because these experiments involve the erosion of an existing bed, rather than flow of slurry over a bed, the equations for fluvial transport in Section 2.3 and 2.4 are applicable, not the pipe flow equations in Section 2.5.

The superficial velocity in the proppant-free region between the bed and the top of the slot was not tabulated by Patankar et al. (2002), but it can be calculated by dividing the volumetric flow rate by $H_1$ and the slot width, 8 mm. Calculations indicate that the Reynolds number of the system at equilibrium was in the range of 1600 to 6000 for most of the experiments. The lowest Reynolds numbers were observed in cases when the aspect ratio was close to one, which is when the critical Reynolds number is lower (Hanks and Ruo, 1956). Overall, the Reynolds number for these experiments was greater than the threshold for the onset of turbulent flow in a slot, except for a few experiments performed with very light-weight beads, as discussed below. Equilibrium velocities were in the range of 25 – 40 cm/s. In experiments with lower viscosity (probably water at around 80°C), Reynolds numbers were around 8000-14,000.

A few experiments were performed with very light-weight beads that were almost neutrally buoyant. These experiments behaved very differently from the other cases. The beads could be transported at a lower velocity than the much denser proppants used in the other experiments. The equilibrium velocity was as low as 7 cm/s, and the flow was laminar in some of the experiments.

Patankar et al. (2002) collected series of measurements where all factors were held equal except volumetric injection rate. In these cases, the critical velocity was fairly consistent as volumetric rate increased (except for the case with the light-weight beads, when velocity increased with volumetric injection rate). This indicates that in each case, flow initiated at a threshold velocity.

The threshold velocity for the onset of bed load transport can be predicted from the Shields number (Equations 3, 5, and 6). For the values of $Re*$ observed in the experiments, the threshold Shields number was in the range of 0.047, the value used in the original Meyer-Peter and Müller (1948) correlation (Miller et al., 1977). The critical velocity can be calculated according to Equation 6, and the equilibrium value of $H_l$ can be calculated from $V_c$, the width, and the volumetric flow rate.

Figure 2 shows a cross-plot between the predicted values $H_l$ and the measured data. The $R^2$ is 0.75. The value of $f_D$ was taken to be 0.04 (discussed in Section 2.1), and the value of $N_{Sh,c}$ was taken to be 0.047. These values were chosen as *a priori* estimates based on the conditions of the experiments and were not varied to minimize to misfit between predictions and observations. Varying these parameters within reasonable ranges has only a minor effect on the fit. If regression had been performed, the fit could have been modestly improved.

The five measurements taken with the light weight beads are not shown in the figure because the predictions deviate strongly from the observations at higher values of $H_l$ (corresponding to higher volumetric flow rate). This is not unexpected because the Shields parameter correlation was developed for application to particles with density in the range of natural sediment, not light weight particles that are almost neutrally buoyant. The results with light-weight particles were unusual because they indicated increasing threshold velocity as volumetric flow rate was increased, in contrast to the other observations.

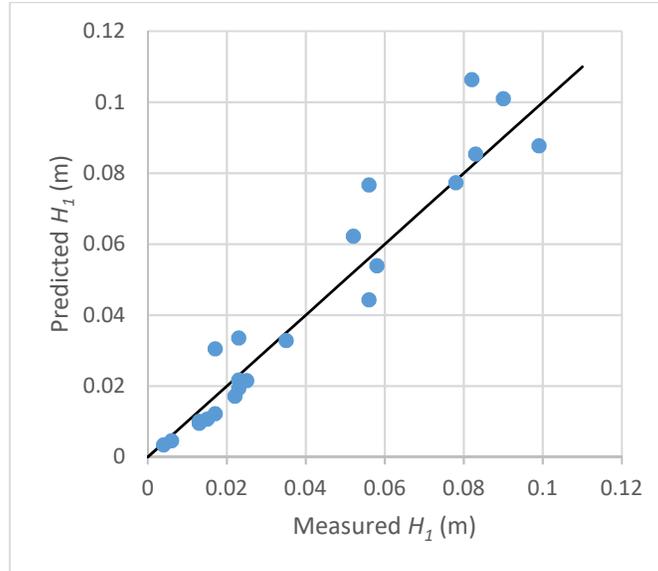

Figure 2: Comparison of measured and predicted $H_1$ values from the Patankar et al. (2002) experiments involving injection of proppant-free fluid over a bed

### 3.2 Comparison with experimental proppant erosion rates from Patankar et al. (2002)

Woodworth and Miskimins (2007) provided additional details about the Patankar et al. (2002) experiments. They noted that when proppant-free fluid was injected into the slot (which was already full of proppant), there was a brief period of relatively rapid erosion, followed by a period in which the erosion rate was on the order of 0.1 – 0.2 cm/min (Figure 7 from Woodworth and Miskimins, 2007). The early erosion was evidently caused by upward movement of fluid around a dune that formed due to stimulated turbulence at the flow inlet (Figure 6 from Woodworth and Miskimins, 2007). The subsequent erosion can be predicted by the bed load transport equations for fluvial erosion that are summarized in Section 2.4.

Woodworth and Miskimins did not report the linear velocity during these erosion periods. For analysis, I make the assumption that flow velocity was 50 cm/s, about 50% greater than the critical velocity for proppant erosion, implying a Shields number in the vicinity of 0.1. From Equation 7, this yields a prediction for the mass rate of proppant erosion, 0.00015 kg/s (assuming specific gravity of 2.65 and 40 mesh). The rate of bed height erosion depends on the length of the channel (because particles eroded upstream can be redeposited further along in the stream). Based on the available descriptions of the experiments, a channel length of 1 m is assumed. Assuming a proppant volumetric fraction of 0.6 in the bed, 0.00015 kg/s of proppant erosion corresponds to a height loss rate of 0.07 cm/min, or one cm every 15 minutes, which is in the range of rates reported by Woodworth and Miskimins (2007).

Equation 7 does not predict any scaling of erosion rate with fracture height, and the rate of height loss decreases as the length of the bed increases because as particles are eroded, they are redeposited further downstream back onto the bed. If flow rate were increased to 2 m/s, a very high velocity for field-scale hydraulic fracturing, the predicted erosion rate would increase to 0.02 kg/s, still a negligibly small amount at the field scale. For a fracture bed that is 100 m long, this would imply an average rate of height loss on the order of 1 mm/minute. On the other hand, at the lab scale, with a proppant bed of length 1 m, this would correspond to a height loss rate of 9 cm/min. Therefore, bed height erosion will occur much more rapidly in the laboratory than in the field. These considerations suggest that proppant bed erosion during field-scale hydraulic fracturing is negligible, except very near the wellbore.

### 3.3 Comparison with experimental data on the rate of proppant slurry transport from Patankar et al. (2002)

In the second set of experiments described by Patankar et al. (2002), proppant was injected in a slurry with water over an existing proppant bed. The proppant deposited, decreasing the cross-sectional area for flow, causing an increase in velocity. Eventually, the proppant stopped depositing and an equilibrium bed height was established.

Tables 3 and 4 from Patankar et al. (2002) provide a detailed summary of their experimental setup and results. Their results are summarized in Table 1.

Table 1: Data from Tables 3 and 4 from Patankar et al. (2002) and values of $V_D$ and $H_I$ calculated from Equation 11

| $Q_s$ (cm³/s) | $Q_t$ (cm³/s) | $H_b$ (cm) | $H_I$ (cm) | $V_{avg}$ (m) | $d$ (m) | $R$ | $V_D$ (m/s) calculated from Equation 11 | $H_I$ (cm) predicted from calculated $V_D$ |
|---|---|---|---|---|---|---|---|---|
| 40 | 284.1 | 1.5 | 2.3 | 1.54 | 0.00085 | 1.65 | 1.58 | 2.3 |
| 45.7 | 288.6 | 1.9 | 2.6 | 1.39 | 0.00085 | 1.65 | 1.63 | 2.2 |
| 28.6 | 279 | 1.3 | 2.3 | 1.52 | 0.00085 | 1.65 | 1.44 | 2.4 |
| 11.4 | 261.2 | 0.9 | 2.4 | 1.36 | 0.00085 | 1.65 | 1.17 | 2.8 |
| 11.4 | 324.9 | 0.9 | 3 | 1.35 | 0.00085 | 1.65 | 1.12 | 3.6 |
| 34.3 | 339 | 1.4 | 2.9 | 1.46 | 0.00085 | 1.65 | 1.44 | 2.9 |
| 11.4 | 326.2 | 0.8 | 3.1 | 1.32 | 0.00085 | 1.65 | 1.11 | 3.7 |
| 45.7 | 349.1 | 1.6 | 3 | 1.45 | 0.00085 | 1.65 | 1.54 | 2.8 |
| 40 | 345.3 | 1.5 | 3 | 1.44 | 0.00085 | 1.65 | 1.49 | 2.9 |
| 28.6 | 334.6 | 1.3 | 2.9 | 1.44 | 0.00085 | 1.65 | 1.37 | 3.0 |
| 22.8 | 328.8 | 1.2 | 2.9 | 1.42 | 0.00085 | 1.65 | 1.30 | 3.2 |
| 17.1 | 332.5 | 1.1 | 3.1 | 1.34 | 0.00085 | 1.65 | 1.21 | 3.4 |
| 5.7 | 319.9 | 0.6 | 3.5 | 1.14 | 0.00085 | 1.65 | 0.98 | 4.1 |
| 2.9 | 316.4 | 0.5 | 4.1 | 0.96 | 0.00085 | 1.65 | 0.89 | 4.4 |
| 1.4 | 314.3 | 0.1 | 5.1 | 0.77 | 0.00085 | 1.65 | 0.82 | 4.8 |
| 0.4 | 312 | 0.1 | 5.8 | 0.67 | 0.00085 | 1.65 | 0.74 | 5.3 |
| 22.3 | 328.9 | 1.3 | 2.9 | 1.42 | 0.001 | 1.73 | 1.38 | 3.0 |
| 11.2 | 318.4 | 0.9 | 2.9 | 1.37 | 0.001 | 1.73 | 1.19 | 3.3 |
| 5.6 | 311.6 | 0.7 | 3.3 | 1.18 | 0.001 | 1.73 | 1.05 | 3.7 |
| 2.8 | 312.5 | 0.8 | 4 | 0.98 | 0.001 | 1.73 | 0.94 | 4.1 |
| 1.4 | 312.4 | 0.8 | 4 | 0.98 | 0.001 | 1.73 | 0.87 | 4.5 |
| 0.7 | 311.7 | 0.7 | 4.2 | 0.93 | 0.001 | 1.73 | 0.82 | 4.8 |
| 0.3 | 316.3 | 0.3 | 5.3 | 0.75 | 0.001 | 1.73 | 0.77 | 5.1 |
| 44.7 | 360.1 | 2.1 | 3.1 | 1.45 | 0.001 | 1.73 | 1.63 | 2.8 |
| 39.1 | 353.3 | 1.8 | 3 | 1.47 | 0.001 | 1.73 | 1.58 | 2.8 |
| 33.5 | 342 | 1.7 | 3 | 1.43 | 0.001 | 1.73 | 1.53 | 2.8 |
| 27.9 | 338.4 | 1.7 | 3 | 1.41 | 0.001 | 1.73 | 1.46 | 2.9 |
| 22.3 | 331.4 | 1.3 | 2.9 | 1.43 | 0.001 | 1.73 | 1.38 | 3.0 |
| 16.8 | 328.9 | 1.2 | 3 | 1.37 | 0.001 | 1.73 | 1.29 | 3.2 |
| 11.2 | 329.8 | 1.1 | 3.2 | 1.29 | 0.001 | 1.73 | 1.18 | 3.5 |
| 5.6 | 326.7 | 0.8 | 3.4 | 1.20 | 0.001 | 1.73 | 1.04 | 3.9 |
| 4.2 | 320.2 | 0.6 | 3.5 | 1.14 | 0.001 | 1.73 | 0.99 | 4.0 |
| 11.2 | 191.6 | 0.8 | 2 | 1.20 | 0.001 | 1.73 | 1.33 | 1.8 |
| 16.8 | 197.2 | 0.9 | 2.1 | 1.17 | 0.001 | 1.73 | 1.47 | 1.7 |
| 22.3 | 202.7 | 1.2 | 1.7 | 1.49 | 0.001 | 1.73 | 1.58 | 1.6 |
| 27.9 | 208.3 | 1.3 | 1.9 | 1.37 | 0.001 | 1.73 | 1.67 | 1.6 |
| 33.5 | 213.9 | 1.5 | 1.9 | 1.41 | 0.001 | 1.73 | 1.75 | 1.5 |
| 44.7 | 225.1 | 1.5 | 1.7 | 1.66 | 0.001 | 1.73 | 1.88 | 1.5 |
| 5.6 | 198 | 0.9 | 2.4 | 1.03 | 0.001 | 1.73 | 1.14 | 2.2 |
| 2.8 | 196.5 | 0.7 | 2.8 | 0.88 | 0.001 | 1.73 | 1.01 | 2.4 |
| 1.4 | 195.1 | 0.2 | 3 | 0.81 | 0.001 | 1.73 | 0.92 | 2.7 |
| 0.7 | 196.3 | 0.5 | 3.3 | 0.74 | 0.001 | 1.73 | 0.85 | 2.9 |
| 0.3 | 195.9 | 0.4 | 4.3 | 0.57 | 0.001 | 1.73 | 0.79 | 3.1 |

As $Q_s$ was increased, $V_{avg}$ and $H_b$ tended to increase and $H_l$ tended to decrease. Figure 3 shows $H_b$ versus proppant injection rate per width, $Q_s/W$. At larger values of $Q_s$, $H_l$ was as low as 1.7 cm, corresponding to a flowing slot aspect ratio of 2.125 (width was 8 mm). The largest aspect ratio in an experiment (corresponding to a very low value of $Q_s$) was 7.25.

Figure 4 shows $V_{avg}$ versus proppant injection rate versus width. The Patankar et al. (2002) results are divided based on whether they used low $Q_t$ (around 196 cm$^3$/s), medium $Q_t$ (around 280 cm$^3$/s), or high $Q_t$ (from 312 – 360 cm$^3$/s). The results from Medlin et al. (1985) are also shown (discussed in Section 3.4, below).

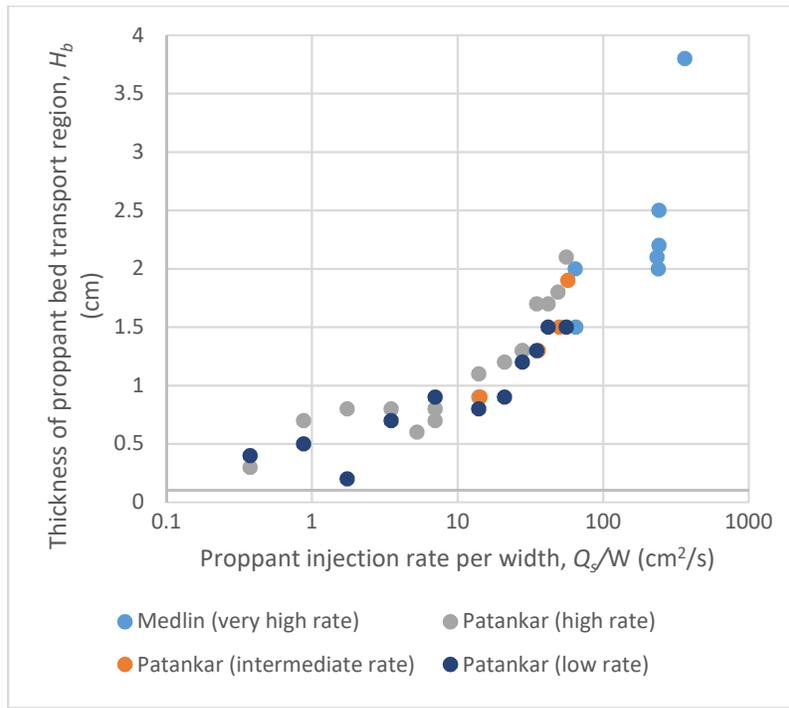

Figure 3: $H_b$ versus $Q_s/W$ from Medlin et al. (1985) and Patankar et al. (2002)

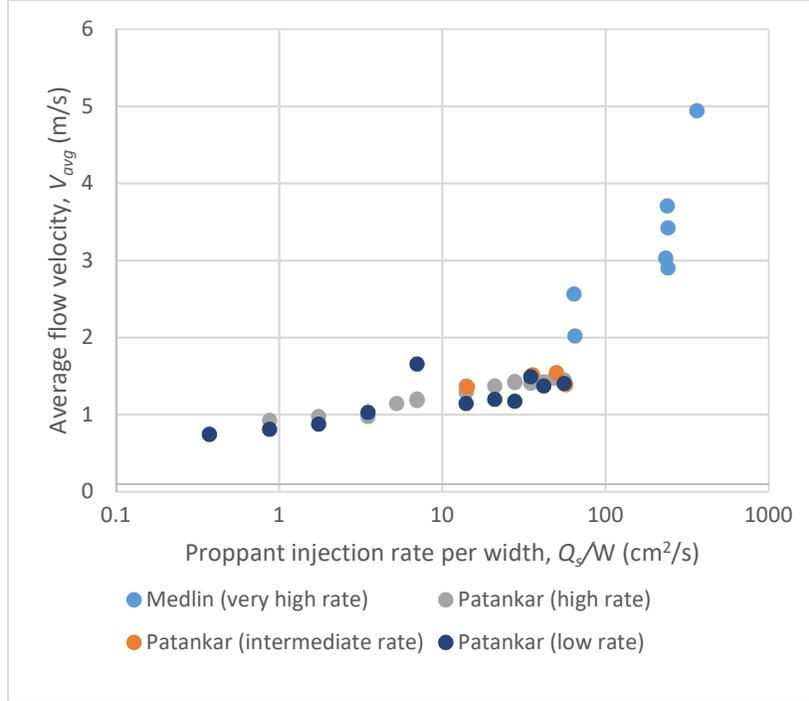

Figure 4: $V_{avg}$ versus $Q_s/W$ from Medlin et al. (1985) and Patankar et al. (2002)

Because the aspect ratio of the flowing region was close to 1.0 in the Patankar et al. (2002) experiments, it is a reasonable hypothesis that Equation 11 should be able to predict $V_{avg}$ (with $D$ equal to the slot width, $W$). If $V_{avg}$ is less than $V_D$, then deposition will occur, increasing $H_l$ and $V_{avg}$, until $V_{avg}$ equals $V_D$ and no further deposition occurs.

At lower values of $Q_s$, corresponding to very dilute proppant concentration, Equation 12 should be able to predict $V_{avg}$. The relatively high aspect ratios (up to 7.25) observed in the dilute experiments Patankar et al. (2002) experiments are likely to be too large for the concentration adjustment in Equation 11 to be valid, but the concentration adjustment in Equation 11 has little effect at dilute concentrations. As concentration is increased, the adjustment in Equation 11 has a greater effect. But at higher concentration, the aspect ratio in the Patankar et al. (2002) experiments was close to 1.0 (as low as 2.1), making the adjustment more likely to be valid.

To test this hypothesis, Equation 11 (with $D$ equal to the slot width, $W$) was used to predict observed $V_{avg}$ and $H_l$ under the assumption that the equilibrium flow velocity would be equal to the predicted $V_D$. Once $V_D$ was calculated from Equation 11, $H_l$ was calculated as the total flow rate divided by width and velocity. Figure 5 shows a cross-plot of the predictions and the observations. The $R^2$ value is 0.81. The calculation assumed $f_D$ equal to 0.04 and had a very weak dependence on $f_D$.

The velocity increased as a function of proppant concentration, from around 0.6 m/s at very low concentrations to 1.6 m/s at high concentrations (at a maximum volume fraction of about 0.2). The increase in $V_D$ is consistent with the prediction of Equation 11. For comparison, the predicted $V_D$ from the simpler Durand (1952) equation is 0.57 m/s. The prediction from the Durand (1952) has only a weak dependence on concentration, which is embedded in the $F_L$ parameter. As a result, the Durand (1952) reasonably predicts $V_D$ at dilute concentration, but underpredicts $V_D$ at higher concentration.

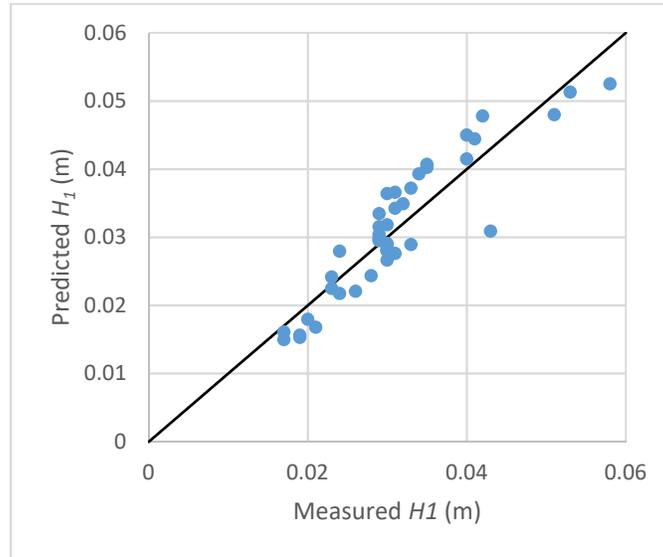

Figure 5: Comparison of measured and predicted $H_1$ values for the Patankar et al. (2002) experiments involving injection of proppant/water slurry

These results indicate that the equilibrium bed height in the Patankar et al. (2002) experiments was controlled by the onset of suspended bed load transport, which was reasonably predicted by correlations for predicting $V_D$ for pipeline flow with $D$ set equal to $W$. However, in these experiments, the aspect ratio of the flowing region of the slot was close to 1.0, especially in the higher concentration experiments. Therefore, these results do not necessarily indicate that Equation 11 can be used to predict $V_D$ in slots with greater aspect ratio. The experiments reviewed in Section 3.4 had higher aspect ratio, and velocity was not well-predicted by Equation 11.

### 3.4 Experimental results from Medlin et al. (1985)

The details of the slot flow experiments by Medlin et al. (1985) are provided in Section 1.2. Medlin et al. (1985) used a slightly wider and longer slot than Patankar et al. (2002), but overall the geometry was similar. Both used water, used a similar range of proppant concentrations, and injected proppant of similar size and density. However, Medlin et al. (1985) used much higher total flow rates, ranging from 1060 cm³/s to 3185 cm³/s, compared with 190 to 360 cm³ by Patankar et al. (2002). Also, Medlin et al. (1985) used higher proppant injection rates, as high as 346 cm³/s, compared to a maximum of 45.7 cm³/s used by Patankar et al. (2002).

Medlin et al. (1985) used an optical absorption method to generate vertical cross-sections of proppant concentration. The full results from all experiments were not tabulated in detail. However, vertical concentration profiles (in regions far from the inlet, where the viscous transport region has disappeared) are provided in Figures 7, 8, 10a, 10b, 11a, 11b, and 12a from Medlin et al. (1985). The bed load transport region is visible as a region where concentration grades roughly linearly from about 0.4-0.5 to $0 - 0.05$. The experimental conditions, $H_l$, $H_b$, and $V_{avg}$ values from these seven figures are tabulated in Table 2. The $H_b$ values tabulated in Table 2 are taken from annotations placed on the figures by Medlin et al. (1985).

Table 2: Results tabulated from the experimental data of Medlin et al. (1985) and $V_D$ and $Q_{s,max}$ values calculated from Equations 11 and 13

| Figure from Medlin et al. (1985) | $Q_s$ (cm$^3$/s) | $Q_t$ (cm$^3$/s) | $H_b$ (cm) | $H_l$ (cm) | $V_{avg}$ (m/s) | $V_D$ calculated from Equation 11 (m/s) | $Q_{s,max}$ calculated from Equation 13 |
|---|---|---|---|---|---|---|---|
| 7 | 228.13 | 2649.71 | 2.0 | 7.5 | 3.71 | 1.37 | 269.28 |
| 8 | 61.23 | 1589.83 | 2.0 | 6.5 | 2.57 | 1.21 | 186.42 |
| 10a | 61.96 | 3179.65 | 1.5 | 16.5 | 2.02 | 1.06 | 146.88 |
| 10b | 230.53 | 3179.65 | 2.2 | 11.5 | 2.90 | 1.41 | 210.74 |
| 11a | 230.53 | 2119.77 | 2.5 | 6.5 | 3.42 | 1.57 | 248.57 |
| 11b | 345.80 | 2119.77 | 3.8 | 4.5 | 4.94 | 1.77 | 359.04 |
| 12a | 223.33 | 1589.83 | 2.1 | 5.5 | 3.03 | 1.69 | 220.32 |

Because of the much greater values of $Q_t$, the values of $H_l$ are much larger in the Medlin et al. (1985) experiments than in the Patankar et al. (2002) experiments. As a result, the aspect ratio in the flowing region ($H_l/W$) ranged from 4.7 to 17.3, significantly greater than in the Patankar et al. (2002) experiments, especially their experiments that used higher proppant concentration. The lowest volumetric fraction reviewed in Table 2 (from their Figure 10a) was only 0.02. However, this cannot be considered "dilute" enough to apply Equation 12 because the total amount of proppant injected was 62 cm$^3$/s, larger than the largest amount injected in any of the Patankar et al. (2002) experiments. As discussed in Section 2.5, Equation 12 is designed for dilute flow in a cylindrical pipe. It could possibly be applied to predict $V_D$ at dilute concentration for flow in a slot, but the concentration required to be sufficiently "dilute" scales inversely with slot height because the proppant will settle and concentrate at the bottom of the slot. The aspect ratio in the Medlin et al. (1985) experiment with injected volume fraction of 0.02 was 17.3.

Because of these considerations, we should not expect Equations 11 and 12 to be able to predict $V_{avg}$ in the Medlin et al. (1985) experiments. Consistent with this expectation, Table 2 shows that the values of $V_D$ calculated from Equation 11 are much less than the observed values of $V_{avg}$. This comparison indicates that the scaling of proppant transport with velocity from the Patankar et al. (2002) experiments breaks down for larger slot height (and aspect ratio) and implies that the Patankar et al. (2002) results cannot be directly extrapolated to the field scale.

The largest proppant injection rate was in Table 2 was 345 cm$^3$/s, or 0.91 kg/s, which corresponded to a $V_{avg}$ of 4.9 m/s. This is an exceptionally high velocity that is unlikely to ever occur in the field, except for very near the well. Despite this extremely high velocity, 0.91 kg/s would be a low rate of proppant transport at the field scale. For injection at 0.116 m$^3$ (50 bpm) into one wing of a biwing fracture, at a volume fraction of 0.045 (1 ppg), the proppant injection rate is 15.9 kg/s.

Medlin et al. (1985) reported that in experiments where the bed height exceeded about 2 cm, equilibrium could not be established. Instead, the bed tended to gradually increase in height until the slot screened out. Therefore, only the observations from Figures 8 and 10a in Table 2 from Medlin et al. (1985) can be considered "equilibrium" values. In the other observations, $H_1$ was continuing to decrease and $V_{avg}$ was continuing to increase over time. Because these observations correspond to a period when net deposition was taking place, the actual net flow rate of proppant through the system (flow in minus flow out) was even lower than the $Q_s$ tabulated in Table 2.

Of the cases reviewed in Table 2, the thickness of the bed load transport region, $H_b$, was no greater than 2.5 cm, except in the case with the highest volume fraction, when it reached 3.8 cm. These values of $H_b$ were only slightly greater than the values observed by Patankar et al. (2002), even though much larger volumes of proppant and water were injected.

It is likely that the value of $H_b$ was limited in the Patankar et al. (2002) and Medlin et al. (1985) experiments because they were performed in a slot. In slot flow, the walls of the slot generate drag that weaken the turbulent eddies that enable suspended bed load transport. In other types of bed load transport, $H_b$ can be much larger. For example, in turbidity currents, the bed transport region can be hundreds of meters thick (Meiburg and Kneller, 2010). Suspended bed load transport without deposition is possible in large pipelines, which also implies that the bed load transport region is thicker than a few cm.

The different behavior of $H_b$ can be quantified with the Reynolds number, which expresses the tendency for turbulent flow. The Reynolds number in a slot is proportional to the hydraulic diameter (equal to double the width for the case of an infinitely high slot, and equal to the width for a square slot). In slickwater fracturing, the highest possible Reynolds numbers are on the order of $10^4$-$10^5$. In pipe flow, the Reynolds number scales with diameter, and can exceed $10^6$. In streams and rivers, the Reynolds number scales with depth and can exceed $10^7$. In turbidity currents, the Reynolds number scales with the height of the plume and can exceed $10^9$ (Meiburg et al., 2010).

The largest ratio of $H_b$ to $W$ observed by either Patankar et al. (2002) or Medlin et al. (1985) was 4.0, observed at very high linear flow velocity of 4.9 m/s (Figure 11b from Medlin et al., 1985). This may be an unreliable datapoint because it was taken when the top of the bed transport region had reached the top of the slot, flow was unstable, and screenout was imminent (Figure 11b from Medlin et al., 1985). Neglecting this observation, the largest ratio of $H_b$ to $W$ observed by Medlin et al. (1985) was 2.6. The largest ratio observed by Patankar et al. (2002) was also 2.6.

Medlin et al. (1985) used a very long slot, 6.1 m. Before the equilibrium bed height was established along the entire slot length, the bed height varied along the length of the slot as the system moved toward equilibrium. During these periods, they noted that the thickness of the bed load transport region was constant along the length of the slot (in the region after the viscous drag region disappeared), regardless of $H_1$. This further confirms the expectation that bed load transport does not scale with the height of slot.

### 3.5 Experimental results from Kern et al. (1959)

The experimental setup used by Kern et al. (1959) is described in Section 1.1. Kern et al. (1959) did not report the equilibrium bed height, the thickness of the bed transport region, or the total volumetric flow rate used in each of their experiments. They only provided a plot of $V_{avg}$ at equilibrium versus proppant injection rate.

At very low proppant injection rates, they found that the equilibrium velocity was 0.3 – 0.6 m/s, which is in the range that would be expected from Equation 12 for $V_D$ at dilute concentrations. The equilibrium velocity nonlinearly increased as proppant injection rate was increased. Around 1 – 1.5 m/s, Kern et al. (1959) found they could achieve large increases in proppant injection rate with only minimal further increase in velocity. The maximum proppant injection rate was as high as 0.4 kg/s, which was achieved at around 1.5 m/s. These results caused Kern et al. (1959) to be optimistic about the prospects for high rates of bed load transport at the field scale. In contrast, in the Medlin et al. (1985) experiments, proppant injection rates in this range required a much higher velocity, around 3 - 4 m/s.

Because of issues with the experimental setup, the Kern et al. (1959) experiments should be interpreted with caution. The slot used in the experiment was only 0.56 m long. At a velocity of 1.5 m/s, proppant grains could be transported through the slot in only a fraction of a second, not allowing enough time for gravitational settling. It is likely that most of the proppant grains injected in the high rate experiments were transported through the slot by viscous drag, not bed load transport. Also, turbulence is stimulated in the region near the flow inlet due to the jetting of fluid from the flow inlets into the slot (Medlin et al., 1985; Biot and Medlin, 1985; Patankar et al., 2002; Brannon et al., 2005; Woodworth and Miskimins, 2007). This turbulence greatly increases proppant suspension. Because the length of the Kern et al. (1959) slot was only 0.56 m, it is likely that the entire slot was influenced by the turbulence stimulated at the inlet. Because of these problems, the Kern et al. (1959) results cannot be considered to be representative of field conditions. In contrast to Kern et al. (1959), Patankar et al. (2002), Medlin (1985), and Brannon et al. (2005) used much longer slots and performed measurements at a sufficient distance from the inlet, where stimulated turbulence had subsided.

## 4. Discussion

### 4.1 Scaling of bed transport to the field scale

The review of the experimental results from Patankar et al. (2002) and Medlin et al. (1985) suggests that $H_b$ reaches a maximum value that is limited by the fracture aperture, around $2.6W$ (Section 3.4). The maximum amount of bed load transport will occur under the conditions: (1) the bed transport region has reached its maximum height, (2) the superficial velocity of proppant through the bed transport region is equal to the overall fluid superficial velocity in the slot (implying uniform velocity distribution along the flowing height and no slip between the particles and the fluid), (3) and the proppant volume fraction in the bed load transport region is around 0.3. The absolute maximum possible volume fraction is around 0.6, but the Medlin et al. (1985) data indicated that proppant volume fraction decreased linearly with height above the bed, for an average value around 0.3.

Therefore, based on the postulate that the maximum possible value of $H_b$ is roughly equal to $2.6W$, it is possible to write an estimate for the maximum rate of proppant flow through a slot in bed load transport:

$$Q_{s,max} = 2.6W \times 0.3W V_{avg} = 0.8W^2 V_{avg} = \frac{0.8WQ_t}{H_1}. \tag{13}$$

This relation is based on experiments with sand and water and may not hold for more viscous systems or different values of proppant density.

In the Patankar et al. (2002) experiments, the values of $Q_s$ were always lower than $Q_{s,max}$ calculated from Equation 13 (reaching a maximum of about half of $Q_{s,max}$ in the highest $Q_s$ cases). In the Medlin et al. (1985) experiments, $Q_s$ was roughly equal to $Q_{s,max}$ calculated from Equation 13 for the five observations with largest $Q_s$ (the cases when an equilibrium bed height could not be established) and was less than $Q_{s,max}$ for the two observations with lower $Q_s$ and flow velocity. The Medlin et al. (1985) results apparently indicate that once the conditions are satisfied such that $Q_s$ reaches $Q_{s,max}$, the dynamics of flow in the bed transport region cause some degree of proppant deposition to occur over time.

At the field scale, multiple fractures may form and flow will be divided among them. The overall $Q_{s,max}$ is independent of the number of fractures. If more fractures form, $V_{avg}$ will be proportionally lower, but the number of proppants beds available to support transport will be proportionally greater. However, if $V_{avg}$ is lower, the proppant flow rate will be less likely to reach $Q_{s,max}$. In the Medlin et al. (1985) experiments, $Q_s$ did not reach $Q_{s,max}$ until velocity greater than about 3.0 m/s, which would be an extremely high velocity in field scale fracturing. Therefore, bed load transport will be maximized if a single fracture forms.

For injection into a fracture with length much greater than height, the aperture distribution can be written in terms of the plane strain solution for a crack (Perkins and Kern, 1961; Nordgren, 1972; Crouch and Starfield, 1983). Plugging this equation into Equation 13 yields:

$$Q_{s,max} = 0.8Q_t \frac{1-\nu^2}{E} \pi(P - \sigma_n), \tag{14}$$

where $E$ is Young's modulus, $\nu$ is Poisson's ratio, $P$ is the pressure in the fracture, and $\sigma_n$ is the normal stress.

The average flow velocity along the fracture (assuming the flow velocity is zero) is:

$$V_{avg} = \frac{Q_t}{WH} = Q_t \frac{E}{1-\nu^2} \frac{1}{\pi(P-\sigma_n)} \frac{1}{H^2}. \tag{15}$$

Equation 14 assumes that $W$ is constant along the height of the fracture. Actually, the width is greatest at the midline and tapers to zero at the top and bottom. When injection starts, the proppant bed will be near the bottom of the fracture and so aperture will be below average, and Equation 14 will overpredict $Q_{s,max}$. If the proppant fills the fracture sufficiently, the aperture at the top of the bed will eventually exceed $W$, reaching a maximum at the centerline of the fracture where the aperture is double the average. At that point, $Q_{s,max}$ will be four times greater than the prediction in Equation 14 (because $Q_{s,max}$ scales with the square of width).

The situation most favorable for bed load proppant transport is extremely dilute proppant injection at very high rate, a single biwing fracture, very low leakoff, very low modulus, very high net pressure with excellent height confinement (because height growth would tend to decrease net pressure), very low fracture height (to maximize $V_{avg}$), and injection performed for a long duration of time (so that the proppant bed can build up to the center of the fracture, where width is greatest). For example, for injection at 0.116 m$^3$ (50 bpm) into one wing of a biwing fracture, with net pressure equal to 6.0 MPa, $E$ equal to 10 GPa, $\nu$ equal to 0.1, and with the proppant bed at the centerline of the fracture (which increases $Q_{s,max}$ by a factor of four relative to Equation 14), the estimated value of $Q_{s,max}$ is 692 cm$^3$/s. If the proppant is injected at the very dilute volume fraction of 0.011 (0.25 ppg), the total proppant injection rate equals 1500 cm$^3$/s (assuming proppant density of 2650 kg/m$^3$). Therefore, under this narrow set of conditions, bed load transport might be able to sustain a nonnegligible fraction of the injected proppant. However, this calculation requires that $Q_s$ reaches $Q_{s,max}$, and the Medlin et al. (1985) experiments suggest this does not occur until a velocity around 3.0 m/s. According to Equation 15, for $V_{avg}$ to reach 3.0 m/s, the height would need to be less than 4.5 m (or 9 m once the fracture was filled to its midline by proppant). Decreasing Young's modulus and increasing net pressure increases $Q_{s,max}$ by increasing aperture, but it also tends to reduce $V_{avg}$, making it more difficult for $Q_s$ to reach $Q_{s,max}$.

This set of conditions is unlikely to be met in the majority fracturing treatments. Under conditions with $E$ equal to 20 GPa, net pressure equal to 3.0 MPa, the aperture at the top of the proppant bed equal to the average aperture of the fracture, and injection at 0.116 m$^3$ (50 bpm) with a volume fraction of 0.049 (1.0 ppg), $Q_{s,max}$ equals 43 cm$^3$/s, and the proppant injection rate equals 6000 cm$^3$/s.

## 4.2 Applications to field scale hydraulic fracturing design

The correlations provided by Patankar et al. (2002) and Wang et al. (2003) have been used in field scale fracturing simulators (Weng et al., 2011; Shiozawa and McClure, 2016), and have been applied to field scale fracturing design (Woodworth and Miskimins, 2007). However, the discussion in this paper shows that the results from Patankar et al. (2002) do not scale up to field condition. For higher proppant flow rates in slots with high aspect ratio, the minimum deposition velocity ceases to scale with $V_D$, as calculated from Equation 11. The use of the Wang et al. (2003) correlations at the field scale is not recommended.

Mack et al. (2014) proposed using the Shields criterion (Equation 3) to select injection rate in order to ensure that exceed the critical Shields number. They measured the coefficient of reptation for different proppants in order to evaluate which types of proppant will be transported most effectively in bed load. However, saltation and reputation processes are orders of magnitude too slow to be relevant to the field scale (Sections 2.4 and 3.2). The Shields criterion is appropriate for predicting erosion of proppant from an existing proppant bed, not transport of proppant that has not yet settled. Because proppant bed erosion is so slow, pumping clean water sweeps to erode the bed further into the formation (Woodworth and Miskimins, 2007) is not likely to be effective.

### 4.3 Need for further laboratory experiments

There is a need for experiments that vary $Q_s$, $Q_t$, and $W$ over a wider range of conditions. The Patankar et al. (2002) experiments used low values of $Q_s$ and $Q_t$, and the Medlin et al. (1985) experiments used high values of $Q_s$ and $Q_t$. Both sets of experiments used similar values of $W$. This paper compares results between these two different datasets. It is possible that unknown differences in experimental setup caused differences in the observations. It would be ideal to test this paper's conclusions by taking measurements over the full range of $Q_s$ and $Q_t$ used by both Medlin et al. (1985) and Patankar et al. (2002) in a single experimental setup.

Based on the reviewed experiments, it has been postulated that the maximum bed height is around $2.6W$ (Equation 13). It is assumed that the scaling is linear because the Reynolds number for flow in a slot scales linearly with width. Observations across a larger range of width values could test this postulated relationship. Experiments should be performed at sufficiently high flow rate so that the flowing slot aspect ratio, $H_1/W$ is significantly larger than unity. This condition was met in the Medlin et al. (1985) experiments, but not in the Patankar et al. (2002) experiments. $H_1/H_b$ should be significantly larger than unity in order to avoid potential interference from the top of the slot. The constant of proportionality between maximum bed height and $W$ may depend on proppant properties such as diameter and density.

## 5. Conclusions

The equations reviewed in Sections 2.3 – 2.5 are applied to predict results from published laboratory experiments on proppant transport in slot flow. Good fits are found for published data under a variety of conditions without using adjustable parameters to fit the data. Suspended bed load transport of slurry that has not yet settled into a bed (Section 2.5) is the only mechanism that has the potential to be a significant source of bed load transport at the field scale.

Correlations from the pipe flow literature are available to predict the minimum velocity at which deposition does not occur during slurry flow in a cylindrical pipe, $V_D$. These correlations reasonably predict the equilibrium velocity in the Patankar et al. (2002) experiments. In these experiments, the aspect ratio of the flowing region in the slot was near unity, which is why pipe flow equations (designed for aspect ratio of 1.0) were applicable. The pipe flow correlations do not provide accurate predictions of the equilibrium velocity from the Medlin et al. (1985) experiments. The Medlin et al. (1985) experiments used much higher flow rate, and so the flowing region in the slot had a higher aspect ratio. This demonstrates that the scaling of equilibrium velocity and bed height observed in the Patankar et al. (2002) experiments breaks down at higher flow rates and cannot be extrapolated to the field scale (Wang et al., 2003).

Slot height and aspect ratio are critical because the proppant tends to settle and concentrate at the bottom of the slot. If $H_1$ is sufficiently small, the bed transport region can occupy nearly the entire flowing height of the slot (Tables 1 and 2). The experimental results suggest that the maximum possible thickness of the proppant bed load transport region in slot flow is on the order of several cm, even at high velocity. This maximum height is unaffected by the height of the slot. I postulate that the maximum height is proportional to the slot width because the Reynolds number for flow in a slot is proportional to width. However, further experiments are needed to test this relationship. The limited height of the bed transport region causes the maximum rate of bed load transport to be low (relative to the field scale), even at high flow velocity.

Overall, the analysis and review of the literature indicates that bed load transport will be nearly always be negligible at the field scale. Scaling arguments demonstrate why bed load transport is the dominant process in the lab, but viscous drag is the dominant process at the field scale (Section 1.3; Medlin et al., 1985; Biot and Medlin, 1985).

Experiments across a wider range of values for slot width, across a wider range of proppant injection rates, and at sufficiently high values of $H_1/W$ would be useful for clarifying the scaling of bed load transport under different conditions.

## Acknowledgements

Thank you to Egor Dontsov for his insightful comments after reviewing the manuscript.

## List of variables

$C$: solid volumetric fraction, unitless

$C_i$: injection volume fraction, unitless

$D$: pipe diameter, m

$D_h$: hydraulic diameter, m

$d$: particle diameter, m

$E$: Young's modulus, MPa

$F_L$: factor in the Durand (1952) correlation, dimensionless

$f_D$: Darcy friction factor, dimensionless

$g$: gravitational constant, 9.8 m/s$^2$

$H_1$: the height of the region between the top of the slot and the top of the settled bed, m

$H_2$: distance from the top of the slot to the top of the bed load transport region, m

$H_b$: the height of the bed load transport region, m

$H_s$: proppant settling height, m

$L$: characteristic length scale, m

$L_s$: length of particle transport before settling, m

$N_{Sh}$: Shields number, dimensionless

$N_{Sh,c}$: critical Shields number, dimensionless

$Q_b$: volumetric solid flow rate in bed transport, m$^3$/s

$Q_s$: the volumetric solid flow rate, m$^3$/s

$Q_{s,max}$: the maximum possible volumetric solid flow rate, m$^3$/s

$Q_t$: total volumetric flow rate, m$^3$/s

$Q_w$: volumetric flow rate of water, m$^3$/s

$R$: $(\rho_s - \rho_f)/\rho_f$, unitless

$Re$: Reynolds number, dimensionless

$Re^*$: boundary Reynolds number, dimensionless

$u^*$: shear velocity, m/s

$u_D^*$: critical shear deposition velocity in a pipe, m/s

$u_0^*$: critical shear deposition velocity in a pipe at dilute concentration, m/s

$V$: flow velocity, m/s

$V_{avg}$: average superficial flow velocity in the region between the top of the proppant bed and the top of the slot, m/s

$V_b$: superficial velocity of proppant in the bed transport region, m/s

$V_{c,homogeneous}$: threshold velocity for homogeneous flow in a pipe, m/s

$V_D$: deposition velocity in a pipe, m/s

$V_{D,0}$: deposition velocity in a pipe at dilute concentration, m/s

$V_t$: terminal velocity for proppant settling, m/s

$V_h$: horizontal fluid velocity, m/s

$W$: width, m

$\mu$: fluid viscosity, Pa-s

$\rho_f$: fluid density, kg/m$^3$

$\rho_s$: solid density, kg/m$^3$

$\tau_b$: shear stress acting on the bed, Pa